# Nanoscale-SiC doping for enhancing $J_c$ and $H_{c2}$ in the Superconducting $MgB_2$

S. X. Dou*, V. Braccini**, S. Soltanian*, R. Klie***, Y. Zhu***, S. Li****, X. L. Wang*, and D. Larbalestier**

*Institute for Superconducting and Electronic Materials, University of Wollongong, Northfields Ave. Wollongong, NSW 2522, Australia,

**Applied Superconductivity Centre, University of Wisconsin-Madison, USA and

***Brookhaven National Laboratory, Upton, NY, 11973, USA

****Advanced materials Research Centre, Nanyang Technological University, 639798, Singapore

The effect of nanoscale-SiC doping of $MgB_2$ was investigated using transport and magnetic measurements. It was found that there is a clear correlation between the critical temperature $T_c$, the resistivity $\rho$, the residual resistivity ratio, RRR = R(300K)/R(40K), the irreversibility field H* and the alloying state  in the samples. SiC-doping introduced many nano-scale precipitates, provoking an increase of $\rho$(40K) from 1 $\mu\Omega$-cm (RRR = 15) for the clean limit sample to 300 $\mu\Omega$-cm (RRR = 1.75) for the SiC-doped sample, leading to significant enhancement of $H_{c2}$ and H* with only minor effect on $T_c$. EELS analysis revealed a number of nano-scale impurity phases: $Mg_2Si$, MgO, $MgB_4$, $BO_x$, $Si_xB_yO_z$, BC and unreacted SiC in the doped sample. TEM study showed an extensive domain structure of 2-4nm domains induced by SiC doping. The $J_c$ for the 10% nano-SiC doped sample increased substantially at all fields and temperatures compared to the undoped samples, due to the strong increase in $H_{c2}$ and H* produced by SiC doping.

74.25.Sv, 74.25.Ha, 74.25.Fy, 74.70.Ad

## Introduction

The critical current density ($J_c$) in $MgB_2$ has been a central topic for extensive research efforts since superconductivity in this compound was discovered[1]. A number of techniques have been developed and employed to improve the $J_c$ performance in magnetic fields. By using a chemical doping with nano-particle SiC into $MgB_2$ we have achieved a $J_c$ enhancement in high fields by more than one order of magnitude, with only slight reduction in $T_c$[2]. A high density of nano-inclusions and possible substitution of SiC for B in $MgB_2$ was suggested to add effective pinning centers and raising $J_c$(H) over a wide range of temperatures.

Recently, using high field transport measurements, Gurevich et al have reported the achievement of record high upper critical fields ($H_{c2}$) for high resistivity films and untextured bulk polycrystals[3]. They found that enhancements to the resistivity have a strong influence on $H_{c2}$. The observed remarkable $H_{c2}$ enhancement to almost 50T is a consequence of the two-gap superconductivity of $MgB_2$[3], which offers special opportunities for further $H_{c2}$ increase by tuning the impurity scattering. Nanoscale SiC doping indeed introduces large alloying and potentially greatly raises resistivity too. Thus, we expected that transport measurements on SiC-doped samples would provide additional useful information for understanding the pinning mechanisms and upper critical field behavior of alloyed $MgB_2$. In this paper we report on such transport and magnetic measurement evaluations in combination with TEM observations on the nanoscale-SiC doped $MgB_2$.

## Experimental Details

$MgB_2$ pellet samples were prepared by an in-situ reaction method, described in detail previously[2]. Powders of magnesium (99%) and amorphous boron (99%) were well mixed with SiC nanoparticle powder (size of 10nm to 100nm) with the atomic ratio of $MgB_2$ with 10 wt% of SiC addition (sample B) and without (sample A). Pellets 10 mm in diameter and 2 mm in thickness were made under a uniaxial pressure sealed in the Fe tube and then heated at $800^o$C for 30min in flowing high purity Ar, followed by furnace cooling to room temperature. These two Wollongong samples were compared to two Madison samples, one being a clean limit (sample C) and the second being this same sample exposed to Mg vapor (sample D) as described in detail elsewhere[4].

The resistivity versus temperature curves $\rho$(T), were measured in magnetic fields up to 9 Tesla by a four-probe method at a current density of about 1 A/cm$^2$ using a 9 T Physical Property Measurement System (PPMS, Quantum Design). From the resistivity curves, we defined the upper critical field as R(H$_{c2}$)

= 0.9 R(T$_c$). We suppose that this critical field is that of the highest orientation, that is with H parallel to the Mg and B planes. Magnetization was measured from 5 to 30 K using an Oxford 14 T vibrating sample magnetometer (VSM). Bar-shaped Samples of about the same size were cut from the as-sintered pellets to minimize size-dependent effect[5]. Magnetic J$_c$ values were determined from the magnetization hysteresis loops using the appropriate critical state model[6]. An empirical magnetic irreversibility line H$_M^*$ was defined at the field at which J$_c$ falls to 100 A/cm$^2$ [4]. This corresponds we believe to the loss of full-sample connectivity close to H$_{c2}$ perpendicular to the Mg and B planes (H$_{c2}^\perp$), since H$^*$ is ~0.9 H$_{c2}$[7]. High-resolution transmission electron microscope (HRTEM) was employed to characterize the morphology of the samples. Electron energy loss spectroscopy (EELS)[8] was obtained using a JEOL-3000F field emission STEM/TEM, equipped with a Schottky field-emission source operated at 300keV.

## Results and Discussion:

Fig. 1 shows the resistivity curves ρ(T) up to 9 T for the undoped (A) and the SiC-doped (B) samples. The onset T$_c$ of the undoped sample was 37.5 K. For the 10wt% SiC-doped sample, T$_c$ decreased only by 0.6K. By contrast, T$_c$ is depressed to about 22K for C doped MgB$_2$ with nominal stoichiometry of Mg(B$_{0.8}$C$_{0.2}$)$_2$ synthesized from Mg and B$_4$C[9]. These results suggest little if any C substitution for B in the doped sample. It is also clear that the doped sample has stronger superconductivity than the undoped, as is shown explicitly in Fig. 2, where the 90% values of the resistive transitions from Fig. 1 are shown. A further important point is that the nominal resistivities of the two samples are very different, ρ(40K) being 90 μΩcm for the undoped sample and 300 μΩcm for the doped sample. We consider that the 90% transition approximates H$_{c2}^\parallel$. Fig.2 also includes the same data taken on the clean-limit (ρ(40K) = 1 μΩcm) and Mg-exposed sample (ρ(40K) = 18 μΩcm) of Braccini et al. [4]. The clean limit (ρ(40K) = 1 μΩcm) and the undoped samples have H$_{c2}^\parallel$ values of 3.1 and 3.3 T at 30K, while the Mg-exposed (ρ(40K) = 18 μΩcm) and SiC-doped are respectively 5.1 and 4.6 T at 30 K. Since H$_{c2}$ should be directly tied to ρ it is clear percolative normal state effects must be considered to be affecting the derived values of resistivity, which are not directly interpretable as being representative of the scattering in the samples.

Fig. 3 shows the J$_c$(H) curves for the undoped and SiC-doped MgB$_2$ samples at 4.2K, 10K, 20K and 30K. Consistent with its higher H$_{c2}$, the doped sample shows smaller dependence of J$_c$ on magnetic field at all temperatures. At 4.2K and low field both samples attain about 10$^5$ A/cm$^2$, while falling to 100 A/cm$^2$ at 12T and 9T at 10K and 7.4T and 5.6T at 20K for doped and undoped samples, respectively. Both samples show a similar exponential fall-off of J$_c$ with increasing field. Fig. 4 shows that the J$_c$ values for the Wollongong samples are higher than for the two Madison samples, perhaps because the latter samples have about 40% porosity [4]. At 20K, the 10wt% SiC doped sample achieved $10^5 A/cm^2$ at 3T, comparable to that of state-of-the-art Ag/Bi-2223 tapes and an order of magnitude higher than recent state-of-the-art Fe/MgB$_2$ tapes[10]. These results significantly strengthen the position of MgB$_2$ as a competitor for both low and high temperature superconductors.

The irreversibility fields ($H_M*$), derived from the fields at which the magnetic hysteresis loops obtained with the VSM indicate that $J_c = 100 A/cm^2$ are shown in Fig. 5. We believe that $H_M*$ corresponds to about 90% of $H_{c2}^{\perp}$. Doping with SiC significantly improved $H_M*$. For example, the $H_M*$ for the SiC doped sample reached 7.4T, compared to 5.6 for the undoped one, 5.2T for the Mg vapour treated one and 3.8T for the clean limit one at 20K.

To understand the significant enhancement of $J_c$ at higher fields for the nano-SiC doping we compare the resistivity $\rho$ and residual resistivity ratio (RRR) for sample A and B, as shown from the resistivity versus temperature curves reported in Fig. 6. For comparison we list some literature data in Table 1. The highest value of $H*_M$ correlates well to the highest value of resistivity, both being found in the SiC-doped sample for which the $J_c(H)$ characteristics are best too. This correlation is not found for all samples. However, as Rowell[11] has recently shown, many measurements of resistivity do not yield values that are characteristic of the local scattering processes which control $H_{c2}$[12]. The reasons for this is that most samples are less than fully connected, due to the presence of significant sample porosity, cracks or wetting, insulating phases at grain boundaries. In the general case the magnitude of these issue is unknown. In the present case it is in interesting that samples A and D in Table 1 have rather similar $T_c$ (37.2 and 36.9K) and $H*_M$ ((5.5 and 5.0T at 20K). However, their resistivity values are wildly different ($\rho(40K)$ = 90 and 18 $\mu\Omega cm$), quiet unrepresentative of the actual scattering processes which really determine $H_{c2}$ and H*.

| Sample | A: Undoped | B: SiC doped | C: Clean limit | D: Mg vapor treated | Pure bulk (Ref. 13) |
|---|---|---|---|---|---|
| $T_c$ (K) | 37.2 | 36.5 | 39 | 36.9 | 40.2 |
| $\rho$ ($\mu\Omega$-cm) at 40K | 90 | 300 | 1 | 18 | 1 |
| RRR R(300K)/R(40K) | 2.1 | 1.74 | 14.7 | 3 | 19.7 |
| $H_m^*$ (20K) (T ) | 5.6 | 7.4 | 3.9 | 5.2 | 3.8 |

**Table 1. Comparison of $T_c$, resistivity and irreversibility field data for samples A, B, C, D and one literature sample (sintered pellet made from [10]B [13]).**

Recently, Gurevich et al.[3] has reported a record-high $H_{c2}$ (0) = 29T for untextured sample C and $H_{c2}^{\perp}$ (0) = 34T and $H_{c2}^{\perp}$ (0) = 49T for a high resistivity film ($\rho$(40K) = 220 $\mu\Omega$cm) using direct, high-field resistivity measurements[3]. In this study, sample D in Table 1 above was measured after aging for two months, during which time the resistivity, $\rho$, dropped from its original value of 18 to 5 $\mu\Omega$-cm at 40K, while $T_c$ also increased from 36.9 K to 37.7 K, due probably to relaxation of a quenched defect structure. A clean film with a low resistivity of 7 $\mu\Omega$-cm at 40K had $H_{c2}^{\parallel}$ of 29T, in comparison to the 49T of the 220 $\mu\Omega$cm film. It seems likely that the SiC-doped sample with the highest resistivity of 300 $\mu\Omega$-cm will also have a very high $H_{c2}$, at least equal to that of the Mg-vapour-treated sample D.

The TEM examination revealed that there is a number of impurity phases in the form of nano-meter size inclusions inside and in between grains in the nano-SiC doped sample. These impurities include $Mg_2Si$, $MgB_4$ and MgO detected by XRD analysis[2,14], and unreacted SiC, amorphous $BO_x$, $Si_xB_yO_z$ and BC detected by using EELS technique. TEM images show that the grain size of $MgB_2$ is smaller than 100nm. EDX analysis shows that the Mg:Si ratio is identical across the entire sample, indicating that the phase distribution is globally homogeneous. However, nano-scale impurity phases $MgB_4$ and MgO are present within the grains. The presence of oxygen within grains is consistent with the results obtained from the above-mentioned 220 $\mu\Omega$cm thin film with strong pinning where the ratio of Mg:B:O reached 1.0:0.9:07[15]. Fig. 7 is a TEM image showing some unreacted SiC particles and its lattice image. The EELS analysis (convergence angle ($\alpha$) = 13 mrad, collection angle ($\theta_c$) = 18 mrad) shows that this particle is indeed pure SiC without B or any other element in it. The EELS analyses also show other phases present in the SiC doped sample. Fig. 8 (a) shows the EELS spectrum of the $Si_xB_yO_z$ phase with no C. Both the fine-structure of Si and B suggests that the phase is amorphous. Fig. 8(b) is the EELS spectrum of the BC phase. Again, the fine-structure of B suggests that the phase is amorphous. The EELS of amorphous $BO_x$ is shown in Fig. 8(c). These phases are often seen between two grains of $MgB_2$ in the sample.

Based on lattice parameter changes and EDX analysis we suggested that C and Si might substitute into the lattice in an early work[2]. However, recent work on SiC-doped $MgB_2$ single crystal grown under high pressure (30 kbar) and high temperature (1900-1950°C) showed there was only C substitution for B but no Si detected in the crystals. These authors revealed that the C substitution for B is as high as 16%, the highest level of substitution in all the C-doping studies so far[16]. There is a clear trend for the C substitution in $MgB_2$ in the literature data[17,18,19,20]. The higher the sintering temperature is the larger proportion of C is substituted for B in $MgB_2$. As we used relatively lower sintering temperature the C substitution for B is expected to be lower. For the nano-SiC doped sample, Fig. 9 (a) is the Z-contrast image[21,22,23], which shows a typical $MgB_2$ crystal in the [100] orientation. A close up look of the atomic structure of the high-resolution lattice image shows that only the Mg columns are visible (Fig. 9(b)), due to the small scattering amplitude of B[24]. The EELS shows the typical fine-structure for B in $MgB_2$[24], but no C signal can be detected (Fig. 9(c)). It should be noted here that light elements, such as C or B can be detected in concentrations down to 0.2% with 10% accuracy in a matrix such as $MgB_2$. However, it is rather difficult to distinguish a small C signal originating from within the lattice or from surface contamination, as the low signal-to-noise ration of the C core-loss for such low concentrations makes it nearly impossible to distinguish the near-edge fine-structure. Due to the large variety of phases present in the SiC doped sample, it is therefore possible that the C substitution at level 1-3%, which is believed to be quite reasonable from the frame work of literature on C substitution[16-20], can not be readily identified, and more careful analysis is needed.

In addition to the high concentration of the nano-inclusions, there are structural defects observed in the nano-SiC doped sample. Fig. 10 shows a high resolution TEM image of the morphology of the SiC doped sample. The majority of nano-domains have a rectangular shape with a domain size of about 2-4 nm. The domain boundaries trap numerous defects to form nano-defect wells and release the strain caused by the rotation of nanodomains, as reported by Li et al[25]. This nano-domain structure may be the result of a small proportion of C substituted for B. In our recent work, we found that C substitution indeed improved flux pinning, while also depressing $T_c$. It was found that an optimal combination of substitution and addition achieved the best enhancement of flux pinning[26].

## Conclusion

We have demonstrated that nano-scale SiC doping into $MgB_2$ introduces a high concentration of various nano-scale impurity phases which results in a very high resistivity, low residual resistivity ratio and large irreversibility field and upper critical field with modest $T_c$ reduction. The highly dispersed nanoscale precipitates do not appear to cause weak links and may serve as strong pinning centres. The doping with SiC shows enhanced critical current density, irreversibility field and upper critical field in a manner helps make $MgB_2$ potentially competitive with both low and high-$T_c$ superconductors.

## Acknowledgment

The authors thank J. Horvat, M.J. Qin, A. Pan, M. Ionescu, H.K. Liu, S.H. Zhou, L.D. Cooley, M. Tomsic and E.W. Collings for their help in various aspects of this work. The work in Wollongong was supported by the Australian Research Council, Hyper Tech Research Inc OH USA, Alphatech International Ltd, NZ and the University of Wollongong, while that in Madison was supported by NSF through the MRSEC on Nanostructured materials and interfaces.

**Fig. 1. The resistivity vs. temperature in fields up to 9 T for the undoped (A) and SiC-doped (B) samples.**

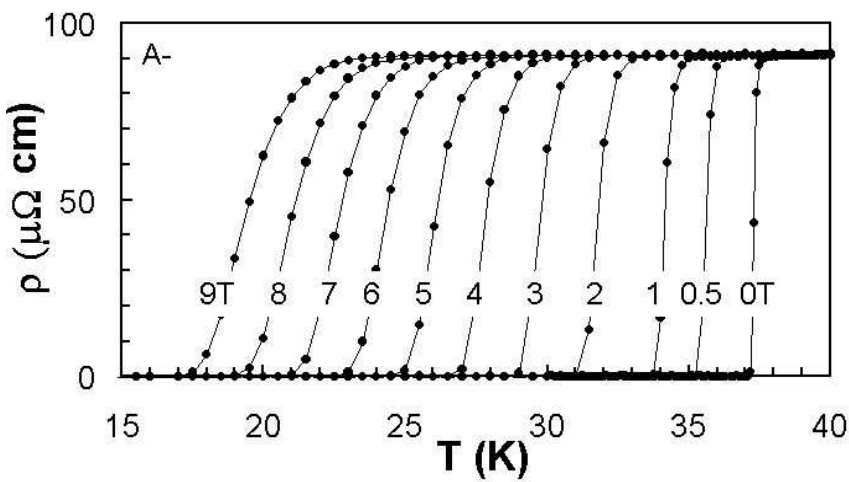

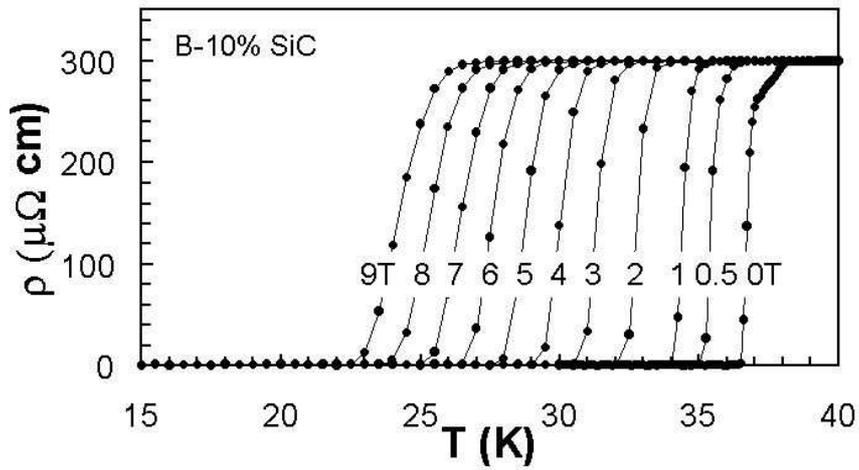

**Fig. 2. The 90% of the resistive transition (upper critical field) as a function of the temperature for the undoped (A), the 10wt% SiC doped sample (B), the clean limit (C) and the Mg vapor treated (D ) samples .**

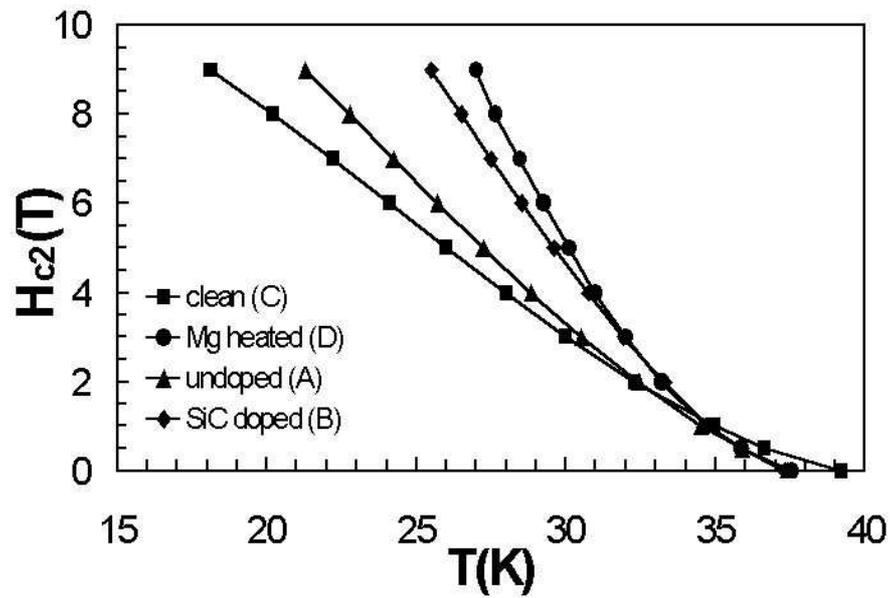

**Fig. 3. The $J_c(H)$ at 4.2, 10, 20 and 30 K for the undoped (A) and SiC doped (B) samples.**

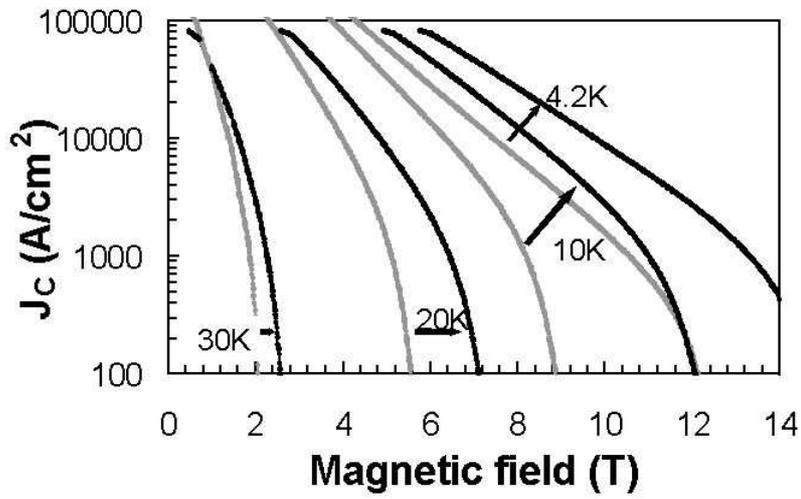

**Fig. 4. A comparison of $J_c(H)$ for the undoped (A), 10wt% SiC doped sample (B), the clean limit (C) and the Mg vapor treated (D ) samples at 4.2 K (a) and 20K (b)**

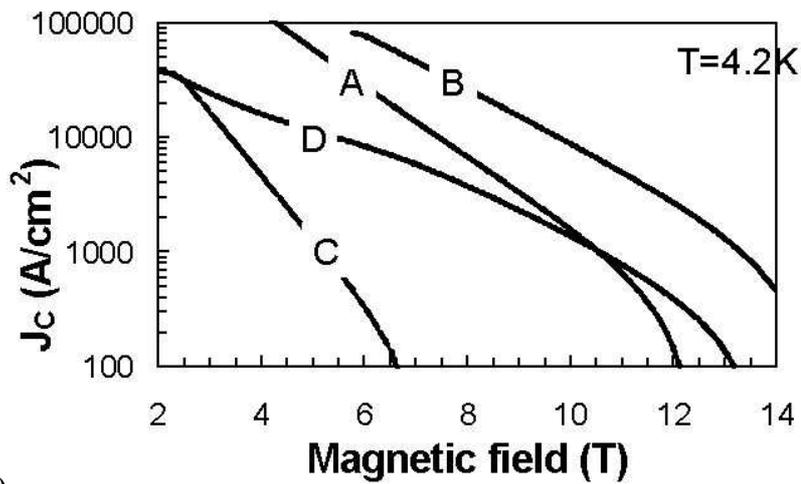

a)

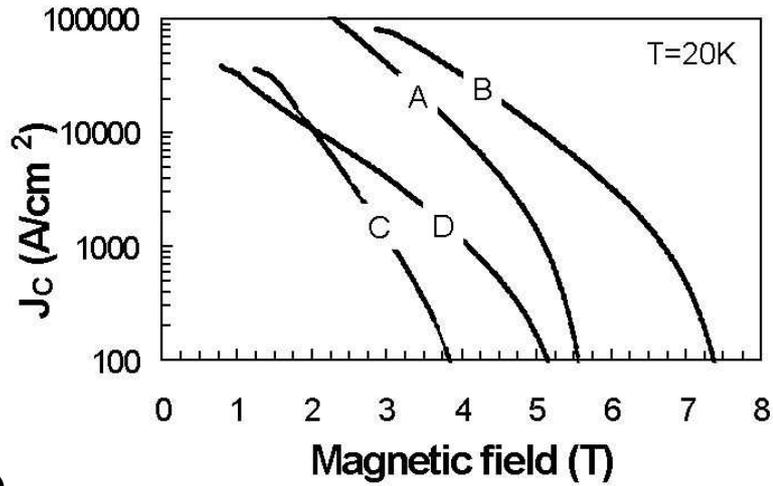

**b)**

**Figure 5: The irreversibility field, $H_M^*$ vs. temperature for the samples A, B, C and D**

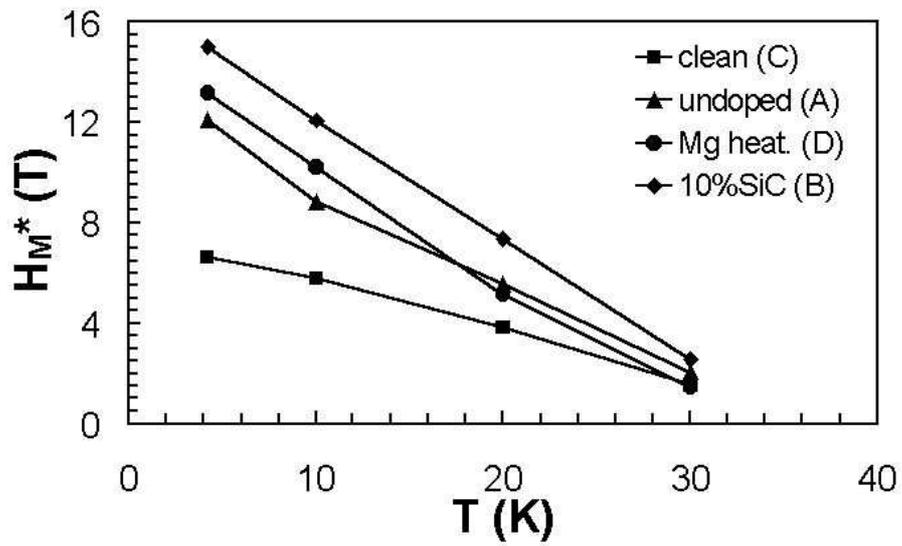

**Figure 6: The resistivity curve as a function of the temperature at zero field for the undoped (A) and SiC-doped (B) samples in the full range 30-300K (a) and near T$_c$ (b)**

a)

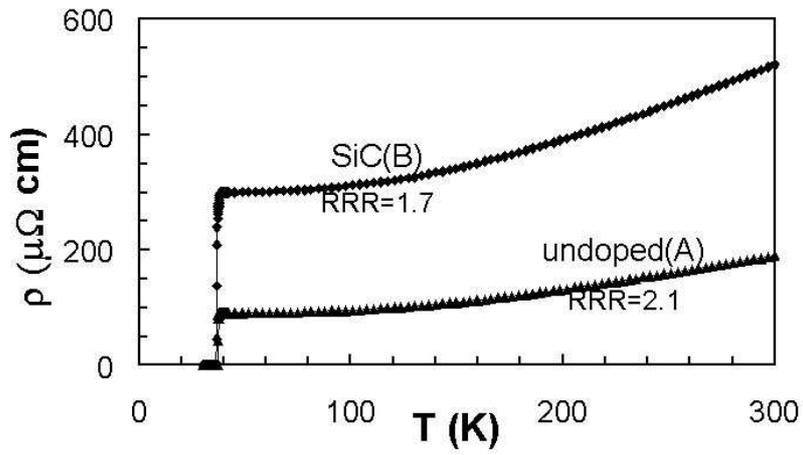

b)

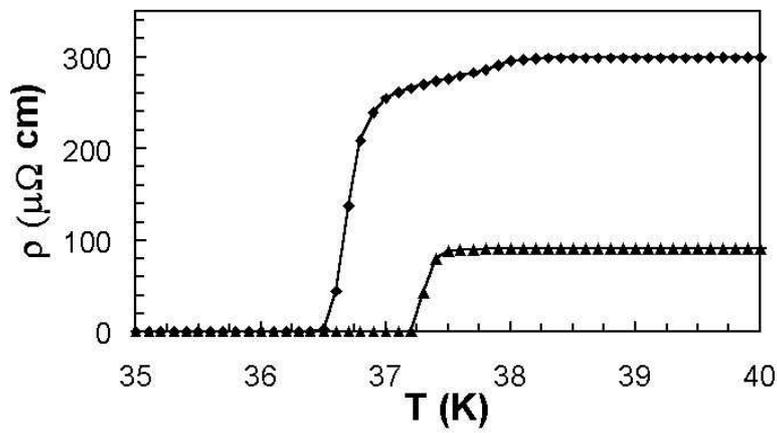

**Figure 7: Conventional TEM image of an unreacted SiC particle; b)
high-resolution TEM image of the bulk of the SiC particle and c)
EELS spectrum clearly showing the Si L- and the C K-edge.**

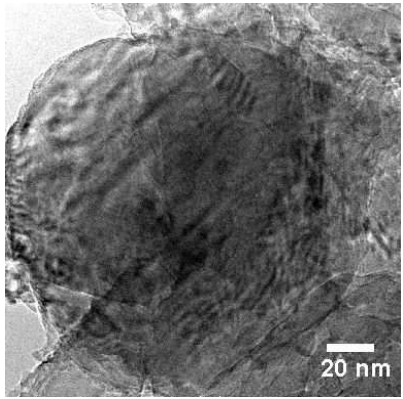

a)

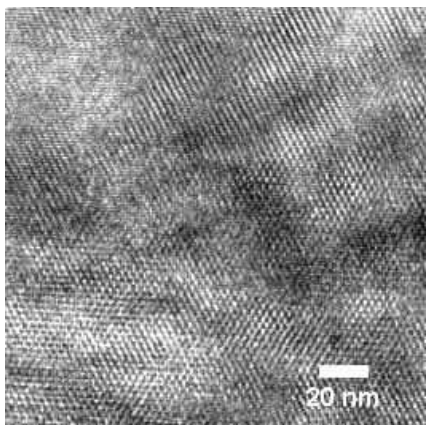

b)

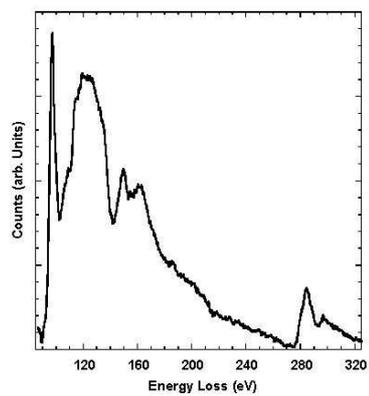

c)

**Figure 8: The EEL spectrum of amorphous a) $BO_x$, b) $Si_xB_yO_z$ and c) BC detected in the SiC doped $MgB_2$.**

a)

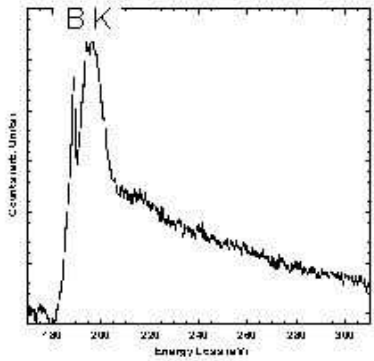

b)

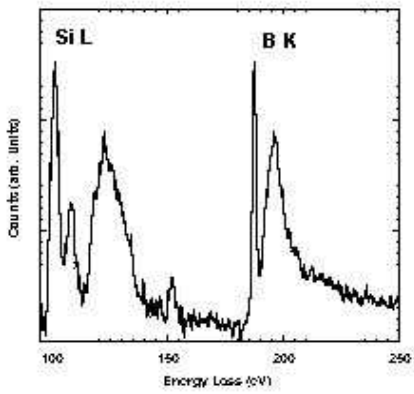

c)

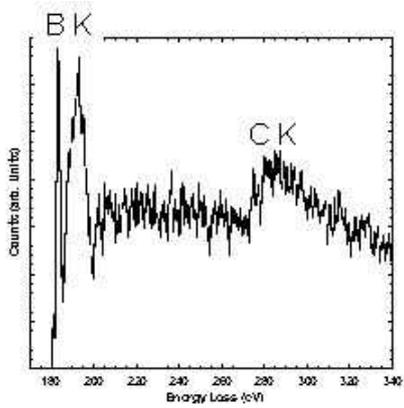

**Figure 9 a) The Z-contrast image of a typical MgB$_2$ grain in the [100] orientation; b) high-resolution Z-contrast image of the bulk of the MgB$_2$ grain showing the Mg columns only and c) EELS spectrum of of the B K-edge from the MgB$_2$ grain.**

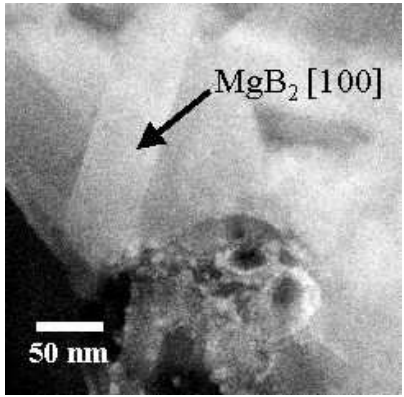

a)

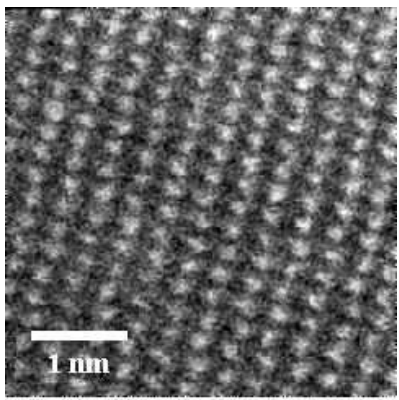

b)

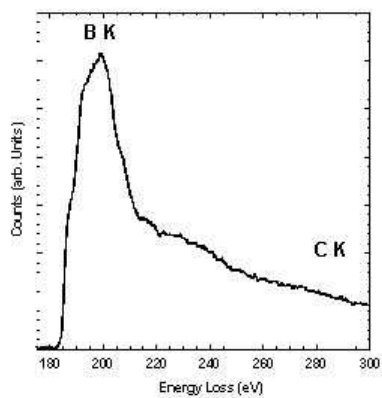

c)

**Figure 10: HRTEM image of nano-domain structure with rectangular nano-wells in the SiC-doped sample (B)**

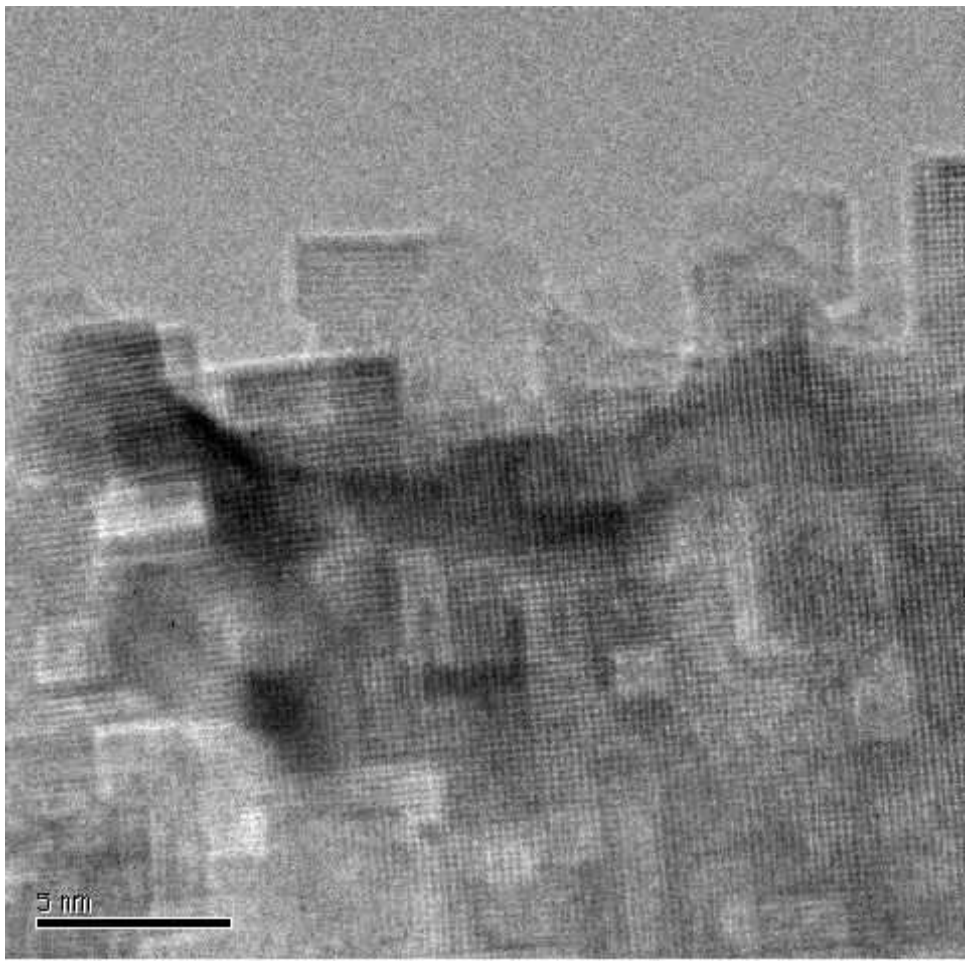

[1] J. Nagamatsu, N. Nakagawa, T. Muranaka, Y. Zenitani and J. Akimitsu, Nature **410**, 63 (2001).

[2] S. X. Dou, S. Soltanian, X. L. Wang, P. Munroe, S. H. Zhou, M. Ionescu, H. K. Liu, and M. Tomsic, Applied Physics Letter **81**, 3419 (2002).

[3] A. Gurevich, S. Patnaik, V. Braccini, K. H. Kim, C. Mielke, X. Song, L. D. Cooley, S. D. Bu, D. M. Kim, J. H. Choi, L. J. Belenky, J. Giencke, M. K. Lee, W. Tian, X. Q. Pan, A. Siri, E. E. Hellstrom, C. B. Eom, D. C. Larbalestier, Cond-mat 0305474.

[4] V. Braccini, L. D. Cooley, S. Patnaik, D. C. Larbalestier, P. Manfrinetti, A. A. Palenzona, and A. S. Siri, Appl. Phys. Lett. **81**, 4577 (2002).

[5] J. Horvat, S. Soltanian, X. L. Wang, and S. X. Dou, cond-mat/0304004.

[6] J. E. J. E. Evetts, *Concise Encyclopedia of Magnetic and Superconducting Materials* (Pergamon, New York, 1992), p.99.

[7] S Patnaik, L. D Cooley, A Gurevich, A. A. Polyanskii, J. Jiang, X. Y. Cai, A. A. Squitieri, M. T. Naus, M. K. Lee, J. H. Choi, L. Belenky, S. D. Bu, J. Letteri, X Song, D. G. Schlom, S. E. Babcock, C. B. Eom, E. E. Hellstrom, and D. C. Larbalestier, Sup. Sci. Tech. **14**, 2001 (315).

[8] R. F. Egerton, *Electron Energy Loss Spectroscopy in the electron microscope* (Plenum, New York, 1986)

[9] R.A Ribeiro, S. Budko, C, Petrovic, and P. C. Canfield, Physica C 384 **384**, 227 (2003).

[10] R. Flukiger, H. L. Suo, N. Mugolino, C. Benedice, P. Toulemonde, and P. Lezza, Physica C **385**, 286 (2003).

[11] J. M. Rowell, Supercond. Sci. Technol. **16**, R17 (2003).

[12] A. Gurevich, Phys. Rev. B **67**, 184515 (2003).

[13] D. K. Finnemore, J. E. Ostenson, S. L. Bud'ko, G. Lapertot, and P. C. Canfield, Phys. Rev. Lett. **86**, 2420 (2001).

[14] S. X. Dou, A. V. Pan, S. Zhou, M. Ionescu, H. K. Liu, and P. R. Munroe,, Supercond. Sci. Technol. **15**, 1 (2002).

[15] C. B. Eom, M. K. Lee, and J. H. Choi, L. J. Belenky, X. Song, L. D. Cooley, M. T. Naus, S. Patnaik, J. Jiang, M. Rikel, A. Polyanskii, A. Gurevich, X. Y. Cai, S. D. Bu, S. E. Babcock, E. E. Hellstrom, D. C. Larbalestier, N. Rogado, K. A. Regan, M. A.

Hayward, T. He, J. S. Slusky, K. Inumaru, M. K. Haas, and R. J. cava, Nature **411**, 558 (2001).

[16] S. M. Kazakov, J. Karpinski, , J. Jun, P. Geiser, N. D. Zhigadlo, P. Puznak, and A. V. Mironov, Cond-mat/0304656.

[17] T. Takenobu, T. Ito, Dam Hieu Chi, K Prassides, and Y. Iwasa,, Phys. Rev. B **64**, 134513 (2001).

[18] I. Maurin, S. Margadonna, K. Prassides, T Takenobu, Y Iwasa, and A. N. Fitch, Chem. Mater. **14**, 3894 (2002).

[19] W. Mickelson, J. Cumings, W. Q. Han, and A Zettl, Phys. Rev. B **65**, 052505 (2002).

[20] Z. H. Cheng and B.G. Shen et al., J. Appl. Phys. **91**, 7125 (2002).

[21] E. M. James and N.D. Browning, Ultramicroscopy **78**, 125 (1999).

[22] N. D. Browning, M. F Chrisholm , and S. J. Pennycook, Nature **366**, 143 (1993).

[23] G. Duscher, N. D. Browning, and S. J. Pennycook, Physica Status Solidi **166**, 327 (1998).

[24] R. F. Klie, J. C. Idrobo, N. D. Browning, K. A. Regan, N. S. Rogado, and R. J. Cava, Appl. Phys. Lett. **79**, 837 (2001).

[25] S. Li, T. White, K. Laursen , T. T. Tan, C. Q. Sun, Z. L. Dong, Y. Li, S. H. Zhou, J. Horvat, and S. X. Dou, Appl. Phys. Lett. **83**, 314 (2003).

[26] S. X. Dou, W. K. Yeoh, J. Horvat, and M Ionescu, submitted.